# Understanding Software in Research: Initial Results from Examining *Nature* and a Call for Collaboration

Udit Nangia
National Center for Supercomputing Applications (NCSA)
University of Illinois Urbana-Champaign
Urbana, USA
udit.nangia2@gmail.com

Daniel S. Katz
NCSA & CS & ECE & iSchool
University of Illinois Urbana-Champaign
Urbana, USA
d.katz@ieee.org

*Abstract*—This lightning talk paper discusses an initial data set that has been gathered to understand the use of software in research, and is intended to spark wider interest in gathering more data. The initial data analyzes three months of articles in the journal *Nature* for software mentions. The wider activity that we seek is a community effort to analyze a wider set of articles, including both a longer timespan of *Nature* articles as well as articles in other journals. Such a collection of data could be used to understand how the role of software has changed over time and how it varies across fields.

*Keywords—software use, software in literature, software citation*

## I. INTRODUCTION

Many scientists, including the authors of this paper, believe that research software is essential and/or central to a large fraction of modern research projects, and that many other projects are dependent on software. We believe this because it is the case in our own research, and when we have asked other researchers, they have told us so. For example, a survey of academic faculty and staff at British universities found that 92% use research software, with 69% saying that their research would not be practical without it [1, 2]. And a recent survey we have performed of members of the (U.S.) National Postdoctoral Association, 95% of respondents use research software, 63% state they could not do their research without research software, 31% could do it but with more effort, and 6% would not find a significant difference in their research without research software [3].

However, one can also study research that has been published to try to understand the role of software in those projects, as has previously been done for a random sample of 90 biology articles by Howison and Bullard [4]. Specifically, this work counted how many papers mentioned software (59 of 90), how many distinct software mentions existed (286 in the 59 papers), how many distinct pieces of software were mentioned (146), and how the software was mentioned (7 different methods, such as citations to a publication, user manual, etc.) A number of others have also performed similar studies, for example, Mayernik et al. [5].

In our work, we have analyzed all papers published in Nature [6] between 1 January and 31 March 2016 for their mentions of software. 80% of these papers mention software, averaging 7 distinct software packages per paper. While Howison and Bullard performed a random sampling, our intent is to be more thorough, in collaboration with others.

## II. METHODOLOGY

The initial work reported here is based on one person's (author Nangia's) reading of the 40 papers that appear in the January – March issues of *Nature*. For each article, Nangia recorded the title, publication date, discipline, and number of mentions of distinct software. For each mention, he recorded the name of the software, the high-level role the software played in the research, if a repository or website was cited, if the software itself was mentioned in the paper's references, and if there was a parenthetical comment after the software name.

## III. RESULTS AND DISCUSSION

Table I shows a summary of the analysis. 32 of the 40 papers examined mention software, and the 32 papers contain 211 mentions of distinct pieces of software, for an average of 6.5 mentions per paper. Across the 32 papers, 173 distinct pieces of software were mentioned, with those mentioned in 4 or more papers shown in Table II.

TABLE I. METRICS ASSOCIATED WITH SURVEY DISTRIBUTION

| | |
|---|---|
| Number of articles | 40 |
| Number of articles that mentioned software | 32 |
| Number of distinct pieces of software mentioned | 211 |
| Total number of distinct pieces of software mentioned | 173 |

a. Distinct within a single article

TABLE II. MOST COMMON SOFTWARE PACKAGES (MENTIONED IN 4 OR MORE PAPERS) AND NUMBER OF PAPERS MENTIONED

| Software Name | Number of Papers Mentioned |
|---|---|
| Pymol | 6 |
| R | 5 |
| Chimera | 4 |
| Coot | 4 |
| Matlab | 4 |
| PHENIX | 4 |

There are some differences apparent between these results and those of Howison and Bullard [4]:

- The fraction of papers that mention software was 66% for Howison and Bullard, but 80% for us.

- Howison and Bullard found each paper that mentioned software mentioned 4.85 distinct pieces of software, while for us, each paper mentioned 6.59 distinct pieces of software.

This leads us to ask if these differences are meaningful. Perhaps they are properties of the fields of the papers, since Howison and Bullard studied biology papers, and we studied papers in a more broad set of disciplines? Or of time, since our papers are more recent than Howison and Bullards, and perhaps software is becoming more important in research? Or perhaps the sample sizes of our studies are too small, and these differences are within the expected errors of a larger study. We suggest a larger collaborative effort could work through these, and additional possibilities, and provide more meaningful data. As a start to this process, our data has been made available [7].

IV. CALL FOR COLLABORATION

At WSSSPE5.2, we will use the lightning talk associated with this paper to propose a collaboration to form a larger community effort to collect and anayze data, similar to what Howison and Bullard and we have previously done.

We will suggest that members of this collaboration can choose a wider set of papers to examine, divide up these papers, and collect the results.

There would then be a number of interesting analyses that could be done, such as examining counts of software vs. time, vs. discipline, vs. journal impact factor, etc. In addition, while we have not done any analysis of the mention styles (e.g., citation to publication, citation to user manual, parenthetical web site of software package,) this could also be studied vs all of the previously mentioned factors.

Of course, in order for this to be meaningful, some initial agreement on how to collect the data is needed, and some checks on agreement after collecting a subset of data will have to be developed. Perhaps there will be a training set, and new data collectors will have to use that training set to obtain the same results as some "gold" standard, or within a set standard of agreement (e.g., Kappa greater than 0.9.)

The mechanism for such a study could be using standard collaborative tools, such as Google docs, sheets, etc., and GitHub. Or it could use a volunteer crowdsourcing platform, such as zooniverse.org. Or a paid platform, such as Amazon's Mechanical Turk.

Finally, AI could be brought in, as Howison and Bullard discussed. Initial human collected data could be used as a training set for machine learning, which could then examine a wider corpus of papers.

V. CONCLUSION

Our initial study of three months of papers in *Nature* provides some evidence that software is important to much of modern science. It confirms earlier work by Howison and Bullard, though also shows some qualitative differences. The significance and potential meaning of these difference is not immediately clear. A larger, collaborative effort to gather similar data will likely lead to a larger data set that can be analyzed to more fully understand the role of software in pubished research.